\documentclass[twocolumn]{jpsj3}
\usepackage{graphicx}
\usepackage{amssymb}
\usepackage{amsmath}
\usepackage{overcite}
\usepackage[nodots,nocompress]{numcompress}
\usepackage{txfonts}

\title{Topological Interaction between Loop Structures in Polymer Networks\\ and the Nonlinear Rubber Elasticity}

\author{Naomi Hirayama\thanks{E-mail address: n-hira@iis.u-tokyo.ac.jp} and Kyoichi Tsurusaki$^{1}$}
\inst{Institute of Industrial Science, University of Tokyo, 4-6-1 Komaba, Meguro, Tokyo, 153-8505, Japan \\
$^{1}$Chemical Engineering Division, Kanagawa Industrial Technology Center, 705-1, Shimoimaizumi, Ebina-city, Kanagawa 243-0435, Japan
}

\abst{We numerically examine the nonlinear rubber elasticity of topologically constrained polymer networks.
We propose a simple and effective model based on Graessley and Pearson's topological model (GP model) for describing the topological effect.
The main point is to take account of a nonequilibrium effect in the synthesis process of the polymer network.
We introduce a new parameter $\gamma$ to describe entropic contributions from the entanglement of polymer loops, which may be determined from the structural characteristics of the sample.
The model is evaluated in the light of experimental data under uniaxial and biaxial deformations.
As a result, our model exhibits uniaxial behaviors which are  common to many elastomers in various deformation regimes such as Mooney-Rivlin's relation in small extension, stress divergence in the elongation limit and the declined stress in compression.
Furthermore, it is also qualitatively consistent with biaxial experiments, which can be explained by few theoretical models.
}

\kword{rubber elasticity, polymer network, topological effect, Mooney-Rivlin's relation, linking probability}

\begin{document}
\maketitle

\section{Introduction}
The topological effect in polymer systems has been attracting considerable attention in fundamental studies as well as in applications~\cite{Tanaka,Iwata1,DE,Fetters,Otto}.
The topologically constrained systems such as ring polymers have been intensively studied~\cite{Tanaka,Iwata1,Fetters,Otto,Toreki,Roovers1,Cho}.
Recent studies~\cite{K1,K2,K3,K4} have evidenced that several familiar polycondensated polymer systems, e.g., polyesters, polycarbonates, poly (ether-sulfon)s and polyamides, also contain large amounts of cyclic polymers.
Such topology of polymer chains have great influences on the systems.
For instance, the relation between the length and the size of an ideal unknotted ring polymer shows a scaling exponent larger than that of an ideal linear polymer.
It can be explained such that a polymer having the trivial knot topology behaves as though it had an excluded volume; that is, the constrained topology statistically results in expansion of the ring polymer~\cite{Shimamura}.
It is also well known that the topological effect is closely related to the osmotic~\cite{Takano,Roovers2,Iwata2} and viscoelastic~\cite{Iwata4,Watanabe,U5} properties.

The topological constraint is expected to influence entropic behaviors of rubber-like materials and cause the complexity and nonlinearities of their elasticity.
A rubber consists of polymer strands cross-linked to each other and the backbones form the network structure.
Strands in the networks cannot self-intersect and intersect with others, and therefore the system is topologically kept unchanged under a deformation.
Such systems demonstrate a nonlinear rubber elasticity~\cite{Treloar,Erman}; i.e., the uniaxial stress does not linearly depend on the strain for large elongation.

The stress-strain relation in rubber-like materials is given by the strain energy density $W$, which is defined as the energy stored in an elastic material under a deformation~\cite{Treloar,Erman}:
\begin{align}
W=\sum_{i,j} C_{i j} (I_1-3)^i (I_2-3)^j.
\label{W}
\end{align}
Here the deformation of the material can be expressed uniquely in terms of the invariants of Green's deformation tensor
\begin{align}
I_1&=\lambda_1^2 + \lambda_2^2 + \lambda_3^2,
\label{I1} \\
I_2&=\lambda_1^2  \lambda_2^2 + \lambda_2^2  \lambda_3^2 + \lambda_3^2  \lambda_1^2,
\label{I2}
\end{align}
where $\lambda_1, \lambda_2$ and $\lambda_3$ are the deformation ratios in the principal axes, and $C_{i j}$ are material constants.
Classical theories assuming an ideal chain network without molecular interactions gives the function $W$ in the form
\begin{align}
W= C_{10} (I_1-3).
\label{classicalW}
\end{align}
It leads to the neo-Hookean stress
\begin{align}
\sigma= 2 C_{1,0} \left( \lambda - \lambda^{-2} \right),
\label{Hook}
\end{align}
where $\lambda$ is the strain ratio in the stretched direction and the incompressibility of a sample is assumed.
This stress (\ref{Hook}) is correct only in the very small extension regime as shown below.

Figure\ \ref{fig1}
\begin{figure}
\begin{center}
\includegraphics[width=7cm]{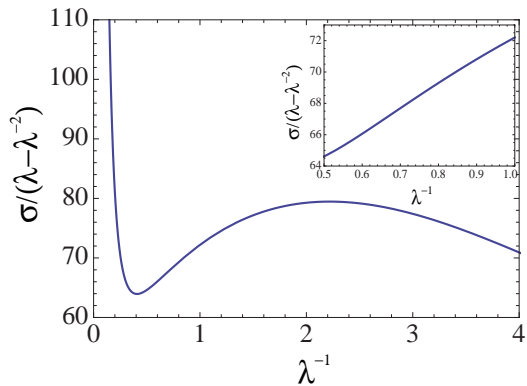}
\caption{The normalized uniaxial stress versus the inverse of the strain curves.
The regimes $\lambda^{-1} < 1$ and $\lambda^{-1} > 1$ of the horizontal axis correspond to elongation and compression, respectively, and the case $\lambda^{-1} =1$ corresponds to the undeformed state.
In the small extension regime ($0.5 \lesssim \lambda < 1$), the curve approximately gives the straight line of Mooney-Rivlin's relation (\ref{MR}) as shown in the inset.
The stress diverges in the elongation limit $\lambda \nearrow \lambda_{\mathrm{c}}$.}
\label{fig1}
\end{center}
\end{figure}
shows the relation between the normalized uniaxial stress versus the inverse of the strain ratio (which we refer to as the Mooney-Rivlin plot hereinafter) obtained from experimental data of end-linking poly (dimethylsiloxane) PDMS network~\cite{U1}.
We calculated the stress from the approximate functional form of $W$ suggested by the experiment~\cite{U1},
\begin{align}
W= C_{10} (I_1-3) + C_{01} (I_2-3) + C_{11} (I_1-3) (I_2-3) 
\notag \\
+ C_{20} (I_1-3)^2 + C_{02} (I_2-3)^2,
\label{5W}
\end{align}
along with the values of $C_{ij}$ reported in Ref.~\cite{U1}.
The neo-Hookean stress (\ref{Hook}) obviously cannot explain the Mooney-Rivlin plot, giving only a flat line.
In the small nonlinear extension regime ($0.5 \lesssim \lambda^{-1} <1$), the stress-strain relation is determined almost exclusively by the first- and second-order terms of the function $W$
\begin{align}
W= C_{10} (I_1-3) + C_{01} (I_2-3).
\end{align}
It leads to Mooney-Rivlin's phenomenological equation~\cite{Erman}
\begin{align}
\frac{\sigma}{\lambda - \lambda^{-2}} = 2C_{10} + 2C_{01}  \lambda^{-1},
\label{MR}
\end{align}
where $C_{10}$ and $C_{01}$ are positive constants.
This would give an increasing line in the Mooney-Rivlin plot in Fig.\ \ref{fig1}.
In the inset, we see that the stress-strain relation indeed obeys Eq.\ (\ref{MR}) in the small extension regime $0.5 \lesssim  \lambda^{-1} < 1$.

However, the second- and higher-order terms are needed for describing the entire deformation regime.
In the large extension regime ($\lambda ^{-1} < 0.5$), the stress $\sigma$ diverges more rapidly with respect to $\lambda$ than Eq.\ (\ref{MR}) predicts and the system reaches the elongation limit $\lambda \nearrow \lambda_{\mathrm{c}}$ at a smaller deformation $\lambda_{\mathrm{c}}$ than the ideal network does.
It can be explained by adding fourth and fifth-order terms in Eq.\ (\ref{5W}).
Further, Eq.\ (\ref{MR}) does not reproduce the behavior in the compression regime ($\lambda^{-1} > 1$), either.
The compressive stress generally gives a declining curve~\cite{Erman,Compression} as shown in Fig.\ \ref{fig1}, contrary to Eq.\ (\ref{MR}).

It is one of our aims in the present paper to explain the above nonlinearities, which correspond to higher-order terms of the function $W$, in terms of the topological effect.
There have been several theories~\cite{Treloar,Erman,Deam,EV,GP,T3} accounting for the nonlinear rubber elasticity from the point of view of the constrained topology of the system.
However, the way that quenched topologies affect the networks is still not fully understood mainly because of their complex structures.
To discuss the effect, we will extend Graessley and Pearson's topological classification model (GP model)~\cite{GP} by introducing a new assumption associated with the nonequilibrium process of the system formation.
As a result, our new model qualitatively reproduces the experimental facts both in extension and compression regimes as reported in Fig.\ \ref{fig6} below.

The paper is organized as follows.
In section 2, we review two topological models: the GP model~\cite{GP} and the simple network model~\cite{T3}.
In section 3, we explain our new model based on the GP model with a new assumption.
In section 4, we numerically evaluate the uniaxial and biaxial stresses and discuss the validity of our model.
Section 5 is devoted to conclusions.

\section{Topological Models}

In the present section, we review two models that partly explain the nonlinear rubber elasticity.
The main purpose of the present paper is to merge the two models and to explain the elasticity in the entire regime.

\subsection{Graessley and Pearson's Topological Model}
\label{GP model}


The GP model~\cite{GP} predicts the elastic contributions of polymer loops having different topologies such as entangled and nonentangled.
The model takes ring-shaped structures, or {\it loops}, as structural elements of a network and employs the following approximations and simplifications:
\begin{enumerate}
\item Loops are randomly distributed in space with the density $\rho_0$.
\item Loop pairs are topologically classified into two cases: mutually entangled with the fraction $P_{\mathrm{L}}$ and not entangled with the fraction $1-P_{\mathrm{L}}$.
\item  Loops are not entangled to themselves, i.e., they have the {\it trivial knot} topology.
\item Distortion of loops are neglected.
\item Networks are affinely deformed by the external force; that is, the vector connecting the centers of mass of the two loops, $\boldsymbol{r}$, is affinely transformed. 
\item The entropic contributions of loop pairs are independent and additive.
\end{enumerate}
%
%
In addition, the GP model assumes the existence of a characteristic distance $S$.
In the following, we will use the radius of gyration of a loop as $S$ and take the normalized vector $\boldsymbol{\zeta} \equiv \boldsymbol{r}/S=(x,y,z)$ instead of $\boldsymbol{r}$.
 
\begin{figure}
\begin{center}
\includegraphics[width=7.5cm,clip]{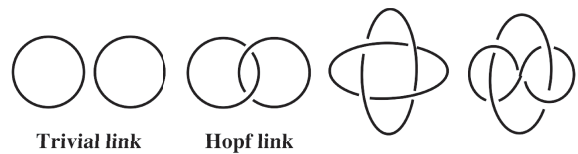}
\caption{{\it The trivial link} and some nontrivial links.
The Hopf link and more complicated link;, i.e., topologies other than the trivial link, are called {\it the nontrivial links}.}
\label{fig2}
\end{center}
\end{figure}

The topology of a loop pair is expressed by a link type.
Some links are depicted in Fig.\ \ref{fig2}.
Under the assumption (ii), the GP model distinguishes only two kinds of topologies, the {\it trivial link} and the {\it nontrivial link}, neglecting detailed topological types of the nontrivial link.
The topologies of the loop pairs are kept unchanged under a deformation; that is, the topological constraint condition exists.

Here, we introduce {\it the linking probability} of a pair of two loops to estimate the entropy of the network.
The linking probability $P_{\mathrm{L}}(\zeta)$ is defined as the following normalized configuration number of entangled loops; $P_{\mathrm{L}}(\zeta) =\Omega_{\mathrm{link}} (\zeta) / (\Omega_{\mathrm{triv}} (\zeta) +\Omega_{\mathrm{link}} (\zeta))$, where $\Omega_{\mathrm{triv}} (\zeta)$ and $\Omega_{\mathrm{link}} (\zeta)$ denote the accessible phase space volume of a loop pair separated in distance $\zeta$ with the trivial link and with the nontrivial link, respectively.
Here we note that each loop has only the trivial knot topology under the assumption (iii).
Thereby, the entropies of a loop pair having trivial and nontrivial link topologies are defined by $s_{\mathrm{triv}} (\zeta)=k_{\mathrm{B}} \ln (1-P_{\mathrm{L}} (\zeta))$ and $s_{\mathrm{link}} (\zeta)=k_{\mathrm{B}} \ln P_{\mathrm{L}} (\zeta)$, respectively, where  $k_{\mathrm{B}}$ is the Boltzmann constant.


After a network reaches the equilibrium state, polymer loops may behave statistically as randomly distributed loops.
Then, the assumption (iv) allows us to calculate $P_{\mathrm{L}}(\zeta)$ both in deformed and undeformed states as the probability for randomly distributed undistorted loops to be entangled.
It can be numerically obtained through the simulation of a random polygon (RP) or a self-avoiding polygon (SAP)~\cite{Hirayama}.
An SAP is modeled by a sequence of $N$ line segments of unit length, namely {\it Kuhn segments}, and hard spherical beads of radius $r_d$, which are placed at vertices to express the excluded volume.
The beads are prevented from intersecting with one another.
The case for $r_d=0$ thus corresponds to an RP.
Hence, $P_{\mathrm{L}}(\zeta)$ can be given by the linking probability of SAPs and RPs.



Under the assumption (v), $\boldsymbol{\zeta}$ changes to $\boldsymbol{\zeta'}=\boldsymbol{\lambda} \cdot \boldsymbol{\zeta}$ if a sample is deformed with the deformation tensor $\boldsymbol{\lambda}=(\lambda_x,\lambda_y,\lambda_z)$.
The entropy change of a nontrivial link is then given as follows: $\Delta s_{\mathrm{link}} (\zeta', \zeta) = s_{\mathrm{link}}(\zeta') - s_{\mathrm{link}}(\zeta) =k_{\mathrm{B}}  \ln (P_{\mathrm{L}}(\zeta')  / P_{\mathrm{L}}( \zeta))$.
The entropy change of a trivial link is similarly given by $\Delta s_{\mathrm{triv}}(\zeta', \zeta)  =k_{\mathrm{B}} \ln \left[ (1-P_{\mathrm{L}}(\zeta')/(1-P_{\mathrm{L}}(\zeta) \right]$.
The GP model then assumes the average entropy change per loop pair of distance $\zeta$ as follows:
\begin{align}
\Delta &s_{\mathrm{loop}} (\zeta', \zeta) = P_{\mathrm{L}}(\zeta) \Delta s_{\mathrm{link}}(\zeta) + (1-P_{\mathrm{L}}(\zeta) ) \Delta s_{\mathrm{triv}}(\zeta)
\nonumber \\
&=k_{\mathrm{B}} \left[P_{\mathrm{L}}(\zeta) \ln \frac{P_{\mathrm{L}}(\zeta')}{P_{\mathrm{L}}(\zeta)} + (1-P_{\mathrm{L}}(\zeta)) \ln \frac{1-P_{\mathrm{L}}(\zeta')}{1-P_{\mathrm{L}}(\zeta)} \right].
\label{delta_s_loop}
\end{align}
Note here that these calculations are done in the framework of equilibrium statistical mechanics.
We will revise this equation in section\ \ref{newmodel} below.

Let us consider that an external force is applied along the $x$-axis.
The separation distances for undeformed and deformed states are given by $\zeta =  |\boldsymbol{\zeta} | =\sqrt{ x^2 + y^2 + z^2 }$ and $\zeta'= | \boldsymbol{\lambda} \cdot \boldsymbol{\zeta} | = \sqrt{ ( \lambda x )^2 + \lambda^{-1} (y^2 + z^2)}$, respectively, where $\lambda=\lambda_1$ is the stretch ratio in the $x$-direction.
(Here we assume the incompressibility of the system and then have $\lambda_2=\lambda_3=\lambda^{-1/2}$.)
Nominal uniaxial stress is then given by $\sigma_{\mathrm{loop}} = - (T/V) ( \partial  \Delta s_{\mathrm{loop}}  / \partial \lambda) $, where $T$ is the temperature.
We thus have the following stress of a loop pair caused by the topological constraint:
\begin{align}
\sigma_{\mathrm{loop}} (\zeta', \zeta) 
= - \frac{k_{\mathrm{B}} T}{V}
 \left[ \frac{P_{\mathrm{L}}(\zeta)}{P_{\mathrm{L}}(\zeta')}
  - \frac{1-P_{\mathrm{L}}(\zeta)}{1-P_{\mathrm{L}}(\zeta')}
\right]
\frac{\partial P_{\mathrm{L}} (\zeta)}{\partial \lambda}.
\label{delta_sigma_loop}
\end{align}

Under the assumption (i), one central loop is paired with $\rho_0 S^3 dx dy dz$ pieces of other loops within a volume element $S^3 dx dy dz$.
Here $\rho_0$ is the loop density defined as $\rho_0=\xi / V$, with the number of loops $\xi$ and the system volume $V$.
Furthermore, under the assumption (vi), the total stress of loop pairs which the central loop participates can be calculated by summing up the contribution from individual pairs all over the system, such that $ \rho_0 S^3 \int_V dx dy dz \, \Delta \sigma_{\mathrm{loop}}$.
Then, the topological stress of the entire system $\Delta \sigma_{\mathrm{GP}}$ can be given by:
\begin{align}
\sigma_{\mathrm{GP}} &=  \frac{ k_{\mathrm{B}} T \rho_0^2 S^3}{2} 
              \int_0^\infty \int_0^\infty \int_0^\infty
                dx \, dy \, dz \,  \notag \\
               &\qquad\qquad\qquad \times \frac{P_{\mathrm{L}}(\zeta') - P_{\mathrm{L}}(\zeta)}{P_{\mathrm{L}}(\zeta') (1-P_{\mathrm{L}}(\zeta'))}
               \frac{ \partial P_{\mathrm{L}}(\zeta')}{\partial \lambda}.
\label{GPsigma} 
\end{align}
The prefactor of $1/2$ compensates for the over-counting of loop pairs.

\begin{figure}
\begin{center}
\includegraphics[width=7cm,clip]{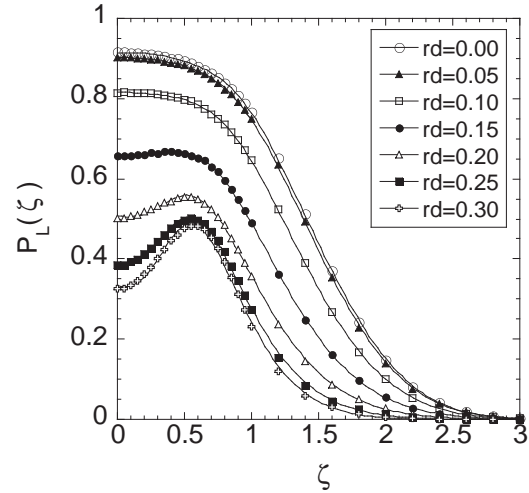}
\caption{Linking probability $P_{\mathrm{L}}(\zeta,N,r_d)$ versus distance $\zeta$ for SAPs of $N=256$ with the following seven values of the excluded volume: $r_d=0.0$ (open circles); $r_d=0.05$ (closed triangles); $r_d=0.1$ (open squares); $r_d=0.15$ (closed circles); $r_d=0.2$ (open triangles); $r_d=0.25$ (closed squares); $r_d=0.3$ (open crosses).
Solid lines are fitting curves given by the formula\ (\ref{P}) with the parameter values reported in Ref.~\cite{Hirayama}.
}
\label{fig3}
\end{center}
\end{figure}

The evaluation of the GP stress requires the precise functional form of $P_{\mathrm{L}}(\zeta)$ since Eq.\ (\ref{GPsigma}) strongly depends on it~\cite{GP}.
The $\zeta$-dependence of $P_{\mathrm{L}}$, however, had remained obscure and the validity of the GP model had not been fully examined.
In the previous study~\cite{Hirayama}, we numerically calculated the linking probability of off-lattice SAPs and proposed the following approximate formula of $P_{\mathrm{L}}$:
\begin{eqnarray}
P_{\mathrm{L}} (\zeta,N,r_d)
= \, \exp(- \alpha \zeta^{\nu_1}) - C \exp(- \beta \zeta^{\nu_2}).
\label{P}
\end{eqnarray}
The fitting parameters $C, \, \alpha, \, \beta, \, \nu_1$ and $\nu_2$ depend on $N$ and $r_d$.
It gave remarkably good fitting curves to the numerical results from the viewpoint of $\chi^2$ values, as shown in Fig.\ \ref{fig3}.

\begin{figure}
\begin{center}
\includegraphics[width=0.7\columnwidth]{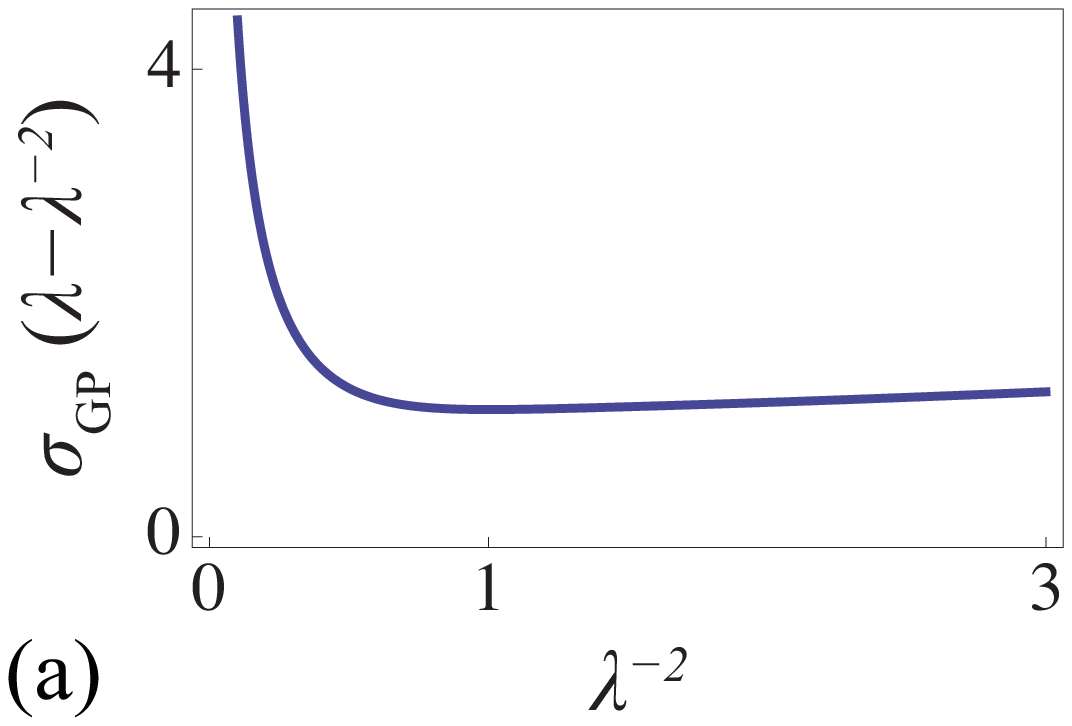}\\
\vspace{\baselineskip}
\includegraphics[width=0.7\columnwidth]{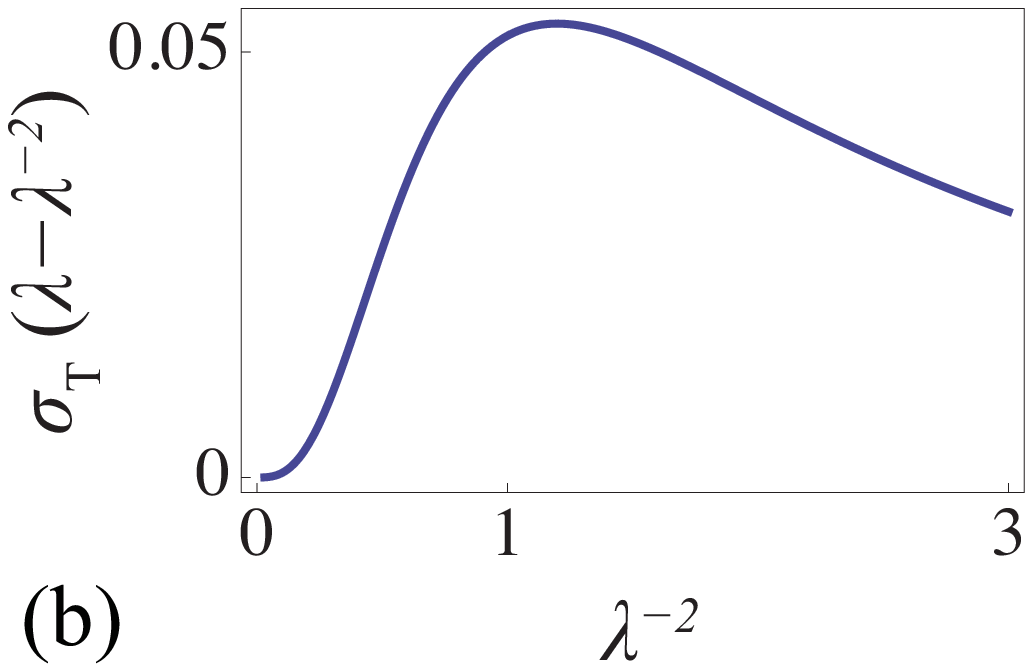} \\
\caption{The Mooney-Rivlin plot for data obtained from (a) the GP model and (b) the simple network model using the function $P(\zeta)$ of Eq.\ (\ref{P}) with the fitting parameters for $r_d=0$ and $N=256$~\cite{Hirayama}.}
\label{fig4}
\end{center}
\end{figure}

The GP stress had not been obtained with a well fitting formula of $P_{\mathrm{L}}$.
To precisely estimate it, we here applied the well fitting formula (\ref{P}) to Eq.\ (\ref{GPsigma}) and obtained the result in Fig.\ \ref{fig4} (a).
The  curve reproduces the stress divergence in the elongation limit $\lambda \nearrow \lambda_{\mathrm{c}}$ as generally observed in real networks.
However, the small extension regime ($0.5 \lesssim \lambda^{-1} < 1$) is contradictory to Mooney-Rivlin's phenomenological law\ (\ref{MR}).
The compressive stress regime ($\lambda^{-1} \ge 1$) is not consistent either with the experiments~\cite{U5,Erman,Compression}, which generally show a decrease of the stress at $\lambda^{-1}$ increases as shown in Fig.\ \ref{fig1}.

\subsection{The Simple Network Model}
\label{SimpleNetwork}
Tsurusaki~\cite{T3} numerically showed that even a network {\it without} entangled loops reproduces Mooney-Rivlin's relation\ (\ref{MR}).
We hereafter refer to it as the simple network model.
The entropy change per loop pair was then given as follows:
\begin{align}
\Delta s_{\mathrm{loop}} (\zeta', \zeta)
=k_{\mathrm{B}} \ln \frac{1-P_{\mathrm{L}}(\zeta')}{1-P_{\mathrm{L}}(\zeta)}.
\end{align}
The first term in the square brackets of Eq. (\ref{delta_s_loop}), giving the contribution of nontrivial link pairs, was omitted here because the network was assumed to have no such loops.

Further, the density of loops paired with the central loop was rewritten by the following function:
\begin{align}
\rho(\zeta)=\rho_0 - V^{-1} e^{-\frac{2}{3} \zeta^2}.
\label{rho}
\end{align}
The second new term in Eq.\ (\ref{rho}) was introduced to accommodate the assumption (i).
It expresses that the central loop has few other loops around itself.
In contrast, the original GP model sets $\rho$ to be the constant $\rho_0=\xi / V$.
It is noticed that $\rho(\zeta)$ increases with $\zeta$ towards $\rho_0$.
The integral of Eq. (\ref{rho}) over the system gives $\xi - 1$, which is the number of loops except the central loop.
Under these new assumptions, the uniaxial stress was given as follows~\cite{T3}:
\begin{align}
\sigma_{\mathrm{T}} =  \frac{ k_{\mathrm{B}} T  \rho_0 S^3}{2} 
              \int_0^\infty \int_0^\infty \int_0^\infty
                dx \, dy \, dz \, \rho (\zeta) 
               \frac{1}{1-P_{\mathrm{L}}(\zeta')}
               \frac{ \partial P_{\mathrm{L}}(\zeta')}{\partial \lambda}.
\label{Tsigma} 
\end{align}

Figure \ref{fig4} (b) shows the elastic behavior of the simple network model obtained with the well fitting function of $P_{\mathrm{L}} (\zeta)$, Eq.\ (\ref{P}).
As a result, the uniaxial stress demonstrated the Mooney-Rivlin elasticity shown in Fig.\ \ref{fig1} for small extensions ($0.5 \lesssim \lambda < 1$).
This suggests that, surprisingly, the additional $C_{01}$ term in Eq.\ (\ref{MR}) results from network structures without entangled loops; i.e., the topological interaction of trivial link pairs, only.
In addition, the compression curve in $\lambda^{-1} <1$ was qualitatively consistent with experimental facts in Fig.\ \ref{fig1}.
However, the simple network model did not reproduce the divergence of the stress in the elongation limit $\lambda \nearrow \lambda_{\mathrm{c}}$.

To summarize so far, the GP model with the well fitting formula of $P_{\mathrm{L}}(\zeta)$, Eq.\ (\ref{P}), can express the elongation limit $\lambda \nearrow \lambda{\mathrm{c}}$, but not the small extension regime or the compression regime.
The simple network model with the new assumption (\ref{rho}), on the other hand, explained Mooney-Rivlin's relation in the small extension regime as well as the compression regime.
Nevertheless, it did not reproduce the elastic behavior under large stretching for $\lambda^{-1} < 0.5$.
These results imply that both models cannot well represent higher-order terms of the phenomenological function $W$ in Eq.\ (\ref{5W}).
The main purpose of the present paper is to explain it in the entire deformation regime by introducing a new assumption in the following section.
By the modification, the model demonstrates good agreement with experiments.

\section{The GP model with a New Assumption}
\label{newmodel}

We improve the GP model by introducing a new assumption in order to describe the entire deformation regime of the rubber elasticity.
Considering possible problems of the GP model, the most significant would be that the fraction of entangled pairs in a system was implicitly postulated to be equal to the linking probability of SAPs.

\begin{figure}
\begin{center}
\includegraphics[width=0.55\columnwidth]{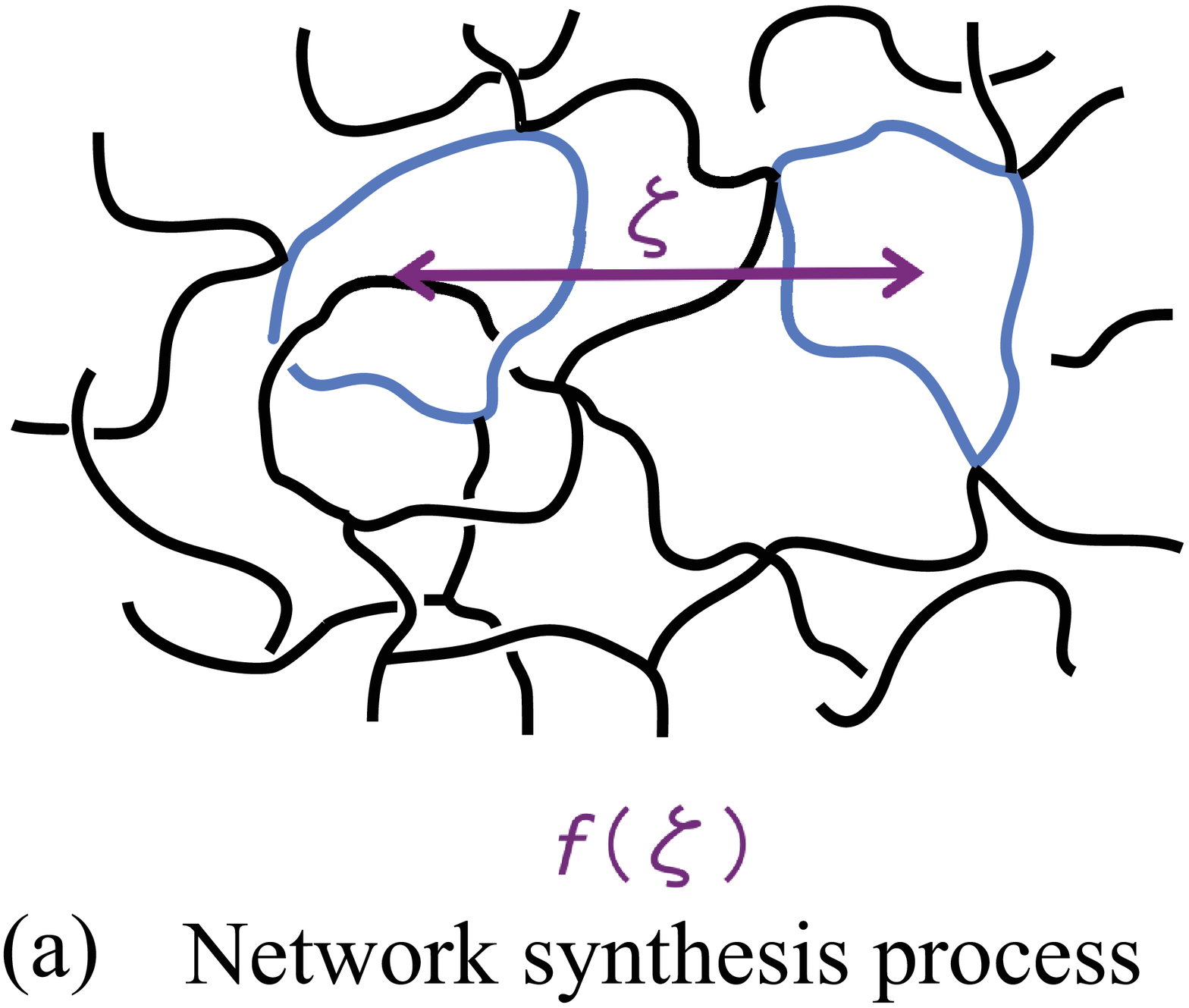} \\

\includegraphics[width=1.0\columnwidth]{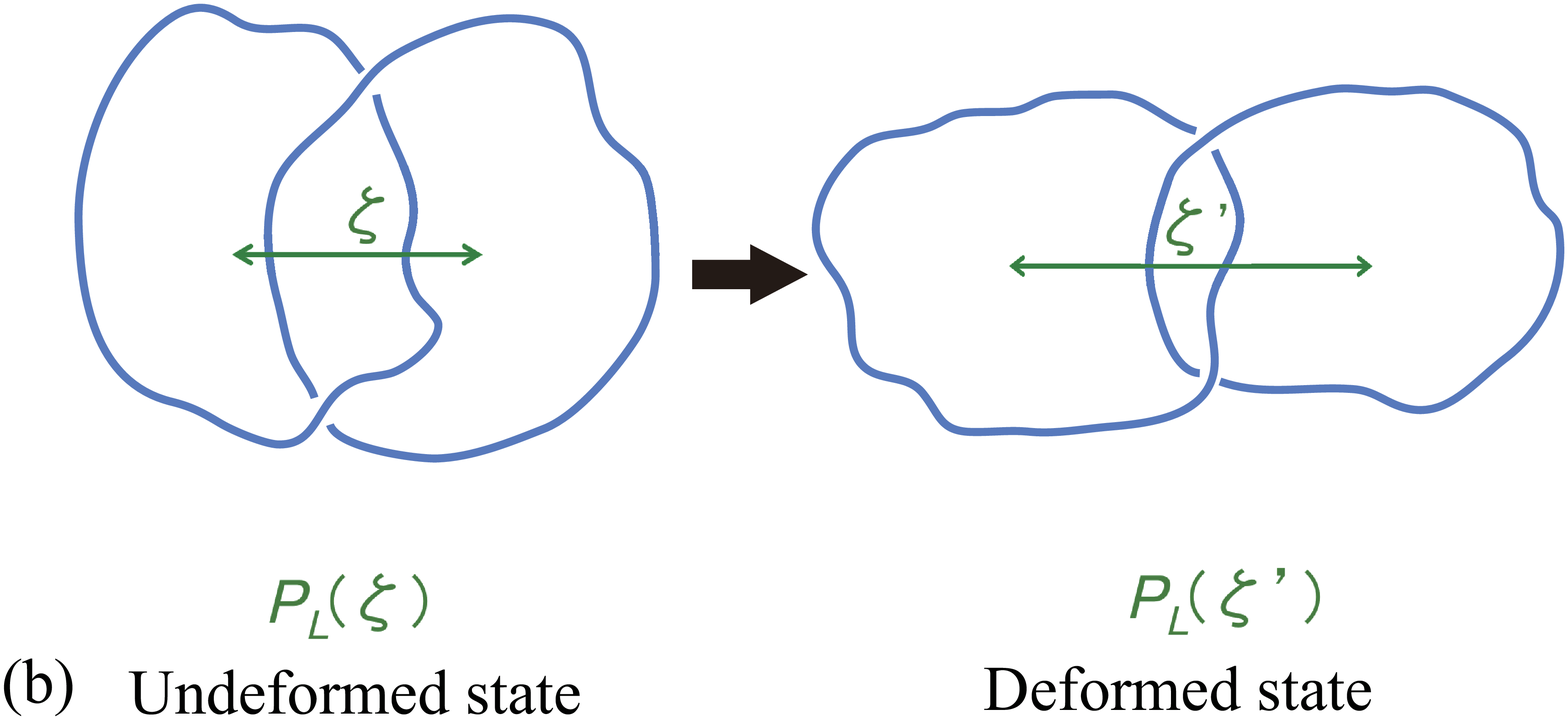} \\
\caption{Linking probabilities of a loop pair (a) in the nonequilibrium synthesis process, $f(\zeta)$ and (b) in the equilibrium state, $P_\mathrm{L} (\zeta)$.
The function  $f(\zeta)$ indicates the probability that two loops of distance $\zeta$ are entangled to each other through the nonequilibrium synthesis process of a network.
On the other hand, $P_\mathrm{L} (\zeta)$ is defined as the linking probability of two equilibrium SAPs.
After the network reaches the equilibrium state, the number of the configurations of the loop pair would be given by  $P_\mathrm{L} (\zeta)$.
Therefore, we express the entropies of a loop pair in the undeformed and deformed systems as $k_B \ln P_\mathrm{L} (\zeta)$ and $k_B \ln P_\mathrm{L} (\zeta')$, respectively.
We thereby obtain the mean entropy change per loop pair by deformation as in Eq.~(\ref{newmodel1}). 
}
\label{fig5}
\end{center}
\end{figure}

A network may be created during a highly nonequilibrium synthesis process.
Then the network entails some loop entanglements with an unknown probability, which should be different from $P_{\mathrm{L}} (\zeta)$ that reflects equilibrium properties.
In order to take the nonequilibrium process into account we now introduce a new function $f(\zeta)$ as the fraction of entangled pairs in the system when they are separated in distance $\zeta$.
Then, the fraction of trivial link pairs is given by $1-f(\zeta)$.
The GP model implicitly assumed that the function $f(\zeta)$ can be replaced by $P_{\mathrm{L}} (\zeta)$ of SAPs~\cite{GP}.
However, such a treatment would be possible only if the synthesis occurred under the equilibrium state.
The topology of loop pairs is in fact determined through the network synthesis process, when the system is under a strongly nonequilibrium condition.
Therefore, the function $f(\zeta)$ must, {\it in principle}, differ from $P_{\mathrm{L}}(\zeta)$.
On the other hand, we assume that the system is under the equilibrium condition during the deformation process.
It corresponds to the situation where the elastic behavior of the network is observed after time enough to reach equilibrium of the synthesized system.
Accordingly, the entropy change of a loop pair is given by the equilibrium function $P_{\mathrm{L}}(\zeta)$ of SAPs, such as $\Delta s_{\mathrm{link}} = \ln \left( P_{\mathrm{L}}(\zeta') / P_{\mathrm{L}}(\zeta) \right)$ for an entangled pair and $\Delta s_{\mathrm{triv}} = \ln \left( 1 - P_{\mathrm{L}}(\zeta') / (1 - P_{\mathrm{L}}(\zeta)) \right)$ for a not-entangled pair.
Figure~\ref{fig5} 
illustrates the distinction between $f(\zeta)$ and $P_\mathrm{L}$.

The contributions from trivial and nontrivial link topologies to the average entropy change, thereby, should be given as follows:
\begin{align}
\Delta &s_{\mathrm{loop}} (\zeta', \zeta) 
= f (\zeta) \Delta s_{\mathrm{link}}(\zeta', \zeta) + (1-f (\zeta) ) \Delta s_{\mathrm{triv}}(\zeta', \zeta)  \notag
\\
&=k_{\mathrm{B}} \left[ f (\zeta) \ln \frac{P_{\mathrm{L}}(\zeta')}{P_{\mathrm{L}}(\zeta)} + (1-f (\zeta)) \ln \frac{1-P_{\mathrm{L}}(\zeta')}{1-P_{\mathrm{L}}(\zeta)} \right].
\label{newmodel1} 
\end{align}
This is the main point of the present paper.
Note again that each entropy change for the trivial and the nontrivial links, $\Delta s_{\mathrm{link}}$ and $\Delta s_{\mathrm{triv}}$, should still be given by the equilibrium function $P_{\mathrm{L}} (\zeta)$ since we consider the elasticity under the equilibrium condition.

Let us recall that the phenomenological Mooney-Rivlin law (\ref{MR}) can be explained by the simple network model without loop entanglements.
It suggests that actual networks possibly have far fewer nontrivial linking than implied by $P_{\mathrm{L}}$, although $f$ and $P_{\mathrm{L}}$ may have similar $\zeta$-dependence.
Therefore, we introduce the following phenomenological assumption; the prefactor $f(\zeta)$ of Eq. (\ref{newmodel1}) is approximately expressed by the form
\begin{align}
f(\zeta)=\gamma P_{\mathrm{L}}(\zeta).
\label{gamma}
\end{align}
Hence, the parameter $\gamma$ is a quantity giving the entropic contribution of the entangled loops.
It should depend on the structural characteristics of a network.
The above consideration leads to the following average entropy change per loop pair:
\begin{align}
&\Delta s_{\mathrm{loop}}(\zeta', \zeta) 
\notag \\
&=k_{\mathrm{B}} \left[ \gamma P_{\mathrm{L}}(\zeta) \ln \frac{P_{\mathrm{L}}(\zeta')}{P_{\mathrm{L}}(\zeta)} + (1-\gamma P_{\mathrm{L}}(\zeta)) \ln \frac{1-P_{\mathrm{L}}(\zeta')}{1-P_{\mathrm{L}}(\zeta)} \right].
\end{align}
The stress contributed by a loop pair is
\begin{align}
\sigma_{\mathrm{loop}} (\zeta', \zeta) 
= - \frac{k_{\mathrm{B}} T}{V}
 \left[ \gamma \frac{P_{\mathrm{L}}(\zeta)}{P_{\mathrm{L}}(\zeta')}
  - \frac{1- \gamma P_{\mathrm{L}}(\zeta)}{1-P_{\mathrm{L}}(\zeta')}
\right]
\frac{\partial P_{\mathrm{L}} (\zeta')}{\partial \lambda}.
\label{new_delta_sigma_loop}
\end{align}

We thus have the uniaxial stress of our new model, with the position-dependent density $\rho(\zeta)$ in Eq. (\ref{rho}), as follows;
\begin{align}
\sigma_{\mathrm{new}} &=  \frac{ k_{\mathrm{B}} T \rho_0 S^3}{2}  
              \int_0^\infty \int_0^\infty \int_0^\infty
                dx \, dy \, dz \,  \rho(\zeta) \notag \\
               &\hspace{7pc} \times \frac{P_{\mathrm{L}}(\zeta') - \gamma P_{\mathrm{L}}(\zeta)}{P_{\mathrm{L}}(\zeta') (1-P_{\mathrm{L}}(\zeta'))}
               \frac{ \partial P_{\mathrm{L}}(\zeta')}{\partial \lambda}.
\label{newsigma} 
\end{align}
We will also use the well fitting formula\ (\ref{P}) for the function $P_\mathrm{L} (\xi)$.

\section{Results and Discussion}
\label{results}

\subsection{Uniaxial Elasticities in Extension and Compression}

\begin{figure}
\begin{center}
\includegraphics[width=7.5cm,clip]{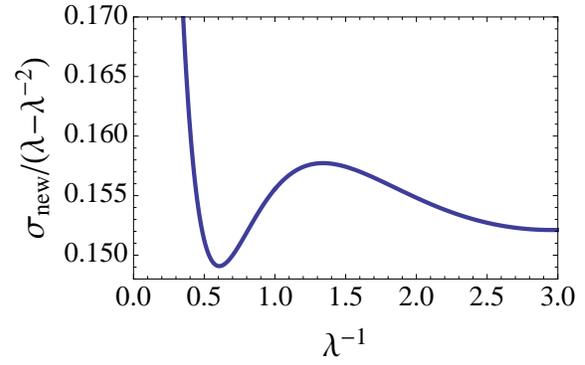} 
\caption{The Mooney-Rivlin plot of the result of our new model\ (\ref{newsigma}) with $\gamma=0.1$.
Here we used Eq. (\ref{P}) in the stress calculation with the fitting parameters for $r_d=0.0$ and $N=256$.}
\label{fig6}
\end{center}
\end{figure}

Figure \ref{fig6}
shows the uniaxial stress of the new model $\sigma_{\mathrm{new}}$ with $\gamma = 0.1$ in the Mooney-Rivlin plot.
The result is in good agreement with the typical experimental data of the nonlinear rubber elasticity throughout the entire deformation regime.
Our new model qualitatively reproduces elastic behavior with higher-order terms of Eq.\ (\ref{5W}) under extension and compression.

The best value of the fitting parameter $\gamma$ can vary depending on the length $N$ and the excluded volume $r_d$ of a loop.
In the case $N=256$ and $r_d=0.0$, the stress successfully fits the empirical form as shown in Fig.\ \ref{fig6} within the values of roughly $0.01 \lesssim \gamma \lesssim 0.1$.
The result suggests that the nontrivial linking rarely occurs in the synthesis process of a real network, contrary to the expectation implied by $P_\mathrm{L}$.
As $\gamma$ increases, the Mooney-Rivlin curve approaches that of the GP model shown in Fig. \ref{fig4} (a), since the number of the nontrivial links increases.

The choice of the parameter $\gamma$ seems to be a feasible task if the network structure can be determined.
It may be, however, exceedingly difficult for real networks because their microscopic topological structures are complex and not directly analyzable.

\subsection{Elastic Contributions from Trivial and Non-Trivial Links}

Both the simple network model and our new model, with a small value of $\gamma$, show the stress-strain relation following Mooney-Rivlin's theory in the small extension regime.
In contrast, the stress divergence under large deformation can be explained only from systems containing the entanglement; such as, the original GP model and our new model with $\gamma \ne 0$.
These results imply that the trivial and nontrivial link topologies govern the system elasticity in different deformation regimes.
Let us now focus on this topic.

\begin{figure}
\begin{center}
\includegraphics[width=7cm,clip]{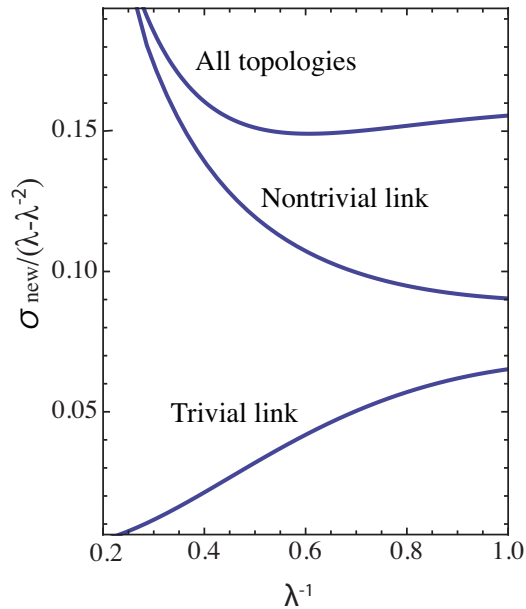}
\caption{The Mooney-Rivlin plots of the stresses of nontrivial and trivial link topologies.
Here we set $r_d=0.0$, $N=256$ and $\gamma=0.1$.}
\label{fig7}
\end{center}
\end{figure}

Figure\ \ref{fig7}
shows the contribution of trivial and nontrivial links to the stress.
Their elastic behaviors are considerably different.
Under small deformations ($0.5 \lesssim \lambda^{-1} < 1$), the change of the normalized stress which is caused by the trivial link exceeds that due to the nontrivial link.
Therefore, the trivial link topology dominates the system elasticity in the small deformation regime and its stress causes Mooney-Rivlin's relation.

Let us explain this behavior in terms of Eq.\ (\ref{new_delta_sigma_loop}).
In Eq.\ (\ref{new_delta_sigma_loop}), the elastic contributions from nontrivial and trivial links are determined by $P_{\mathrm{L}} (\zeta')$.
Under small extensions, loops may be closely distributed to each other and their linking probability $P_{\mathrm{L}}(\zeta')$ should have a large average as seen in the numerical results in Fig.\ \ref{fig3} for small $\zeta$.
In this case, the stress of the trivial links drastically changes with $\lambda$ since the $\lambda$-derivative of the second term in Eq.\ (\ref{new_delta_sigma_loop}) is proportional to $(1-P(\zeta'))^{-2}$ and takes a large value when $P_{\mathrm{L}} (\zeta') \sim 1$; whereas, the stress of the nontrivial links slightly changes because the $\lambda$-derivative of the first term in Eq.\ (\ref{new_delta_sigma_loop}) is proportional to $P(\zeta')^{-2}$.

In the large extension regime ($\lambda^{-1} < 0.5$), on the other hand, the contribution of the trivial links vanishes as seen in Fig.\ \ref{fig7}, and thus the system stress is affected principally by the nontrivial links.
The stress of the nontrivial links diverges in the deformation limit $\lambda_\mathrm{c}$ as shown in Fig.\ \ref{fig7}.
This behavior can be also explained in terms of Eq.\ (\ref{new_delta_sigma_loop}).
The linking probability $P_\mathrm{L} (\zeta)$ for small $r_d$ continuously decreases as $\zeta$ increases as shown in Fig.\ \ref{fig3}.
It makes the first term of Eq.\ (\ref{new_delta_sigma_loop}) dominant under large deformations since the loops tend to be very separated from each other in such a situation.

\begin{figure}
\begin{center}
\includegraphics[width=\columnwidth]{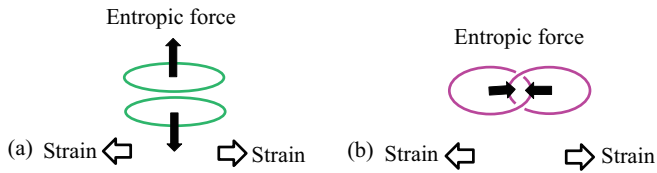}
\caption{Topological interaction between loops having different topologies: (a) the trivial link and (b) the nontrivial link.}
\label{fig8}
\end{center}
\end{figure}

Figure\ \ref{fig8} (a) 
schematically shows how the topological interaction between the loops of a trivial link pair contributes to the stress.
If a trivial link pair is close to each other in the direction perpendicular to a deformation, the entropic restoring force acts on the trivial link, since the loops become closer and interrupt each other's random configurations under the deformation.
This restoring force contributes to the stress in the perpendicular direction because the system is assumed to be imcompressible.
The interaction between such a trivial link pair should be the origin of the Mooney-Rivlin stress expressed as Eq. (\ref{MR}).
Figure\ \ref{fig8} (b), whereas, demonstrates that an entangled loop pair is strongly attracted to each other under a deformation.
Hence, the system stress diverges in the large deformation regime mainly because of the nontrivial link topological interaction.

\subsection{Swollen Effect of the New Model}

\begin{figure}
\begin{center}
\includegraphics[width=7.3cm,clip]{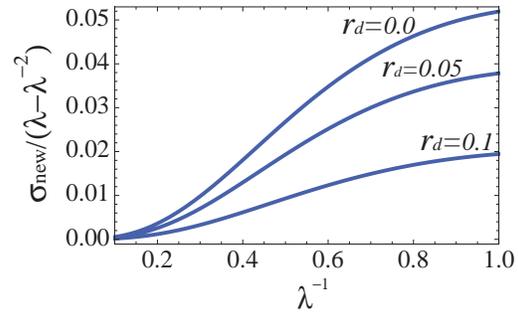} 
\caption{The Mooney-Rivlin plots of our new model with the excluded volume $r_d=0.0$ (top), \, $0.05$ (middle) and $0.1$ (bottom) for the Kuhn segment of an unit length.
We set $N=256$ and $\gamma=0.1$.}
\label{fig9}
\end{center}
\end{figure}

The above considerations are for small $r_d$.
In contrast, highly swollen networks have a large excluded volume and they are strongly influenced by the existence of solvents.
Indeed, it is frequently reported that the additional term $C_{01}$ on Mooney-Rivlin's relation (\ref{MR}) decreases with the volume fraction of polymers~\cite{Erman}.
As demonstrated in Fig.\ \ref{fig9}, our new model is consistent with such experimental facts; the gradient of the Mooney-Rivlin curve decreases with $r_d$ increases in the small deformation regime ($0.5 \lesssim \lambda < 1$).
We can understand the swelling effect from the $r_d$-dependence of $P_{\mathrm{L}}$.
The excluded volume $r_d$ drastically affects $P_{\mathrm{L}}(\zeta)$ in Fig.\ \ref{fig3}.
When $r_d \ge 0.15$, $P_{\mathrm{L}}(\zeta)$ takes rather small values even for $\zeta=0$.
Hence, the topological effect of the trivial link topology should practically vanish since the second term of Eq.\ (\ref{new_delta_sigma_loop}) is relatively negligible over all the deformation regimes.
%

\subsection{Biaxial Stress-Strain Relation}
\begin{figure}
\begin{center}
\includegraphics[width=5cm,clip]{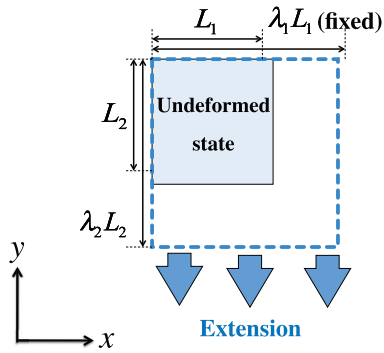} 
\caption{The sample is elongated in the $y$-direction with the nominal strain ratio $\lambda_2$ when the strain ratio in the $x$-direction is fixed at $\lambda_1$.
Nominal stresses along the $x$- and $y$-directions are denoted by $\sigma_1$ and $\sigma_2$, respectively.}
\label{fig10}
\end{center}
\end{figure}

\begin{figure}
\begin{center}
\begin{tabular}{c}
\resizebox{63mm}{!}{\includegraphics{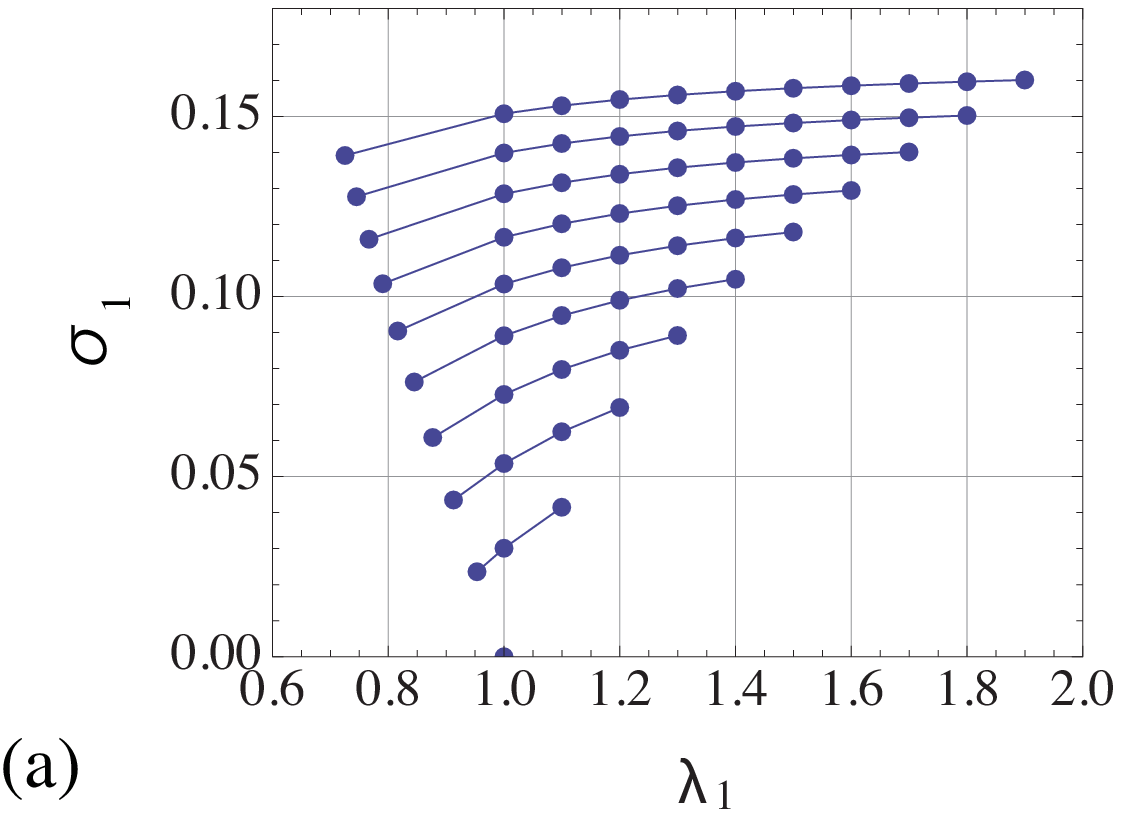}} \\
\resizebox{63mm}{!}{\includegraphics{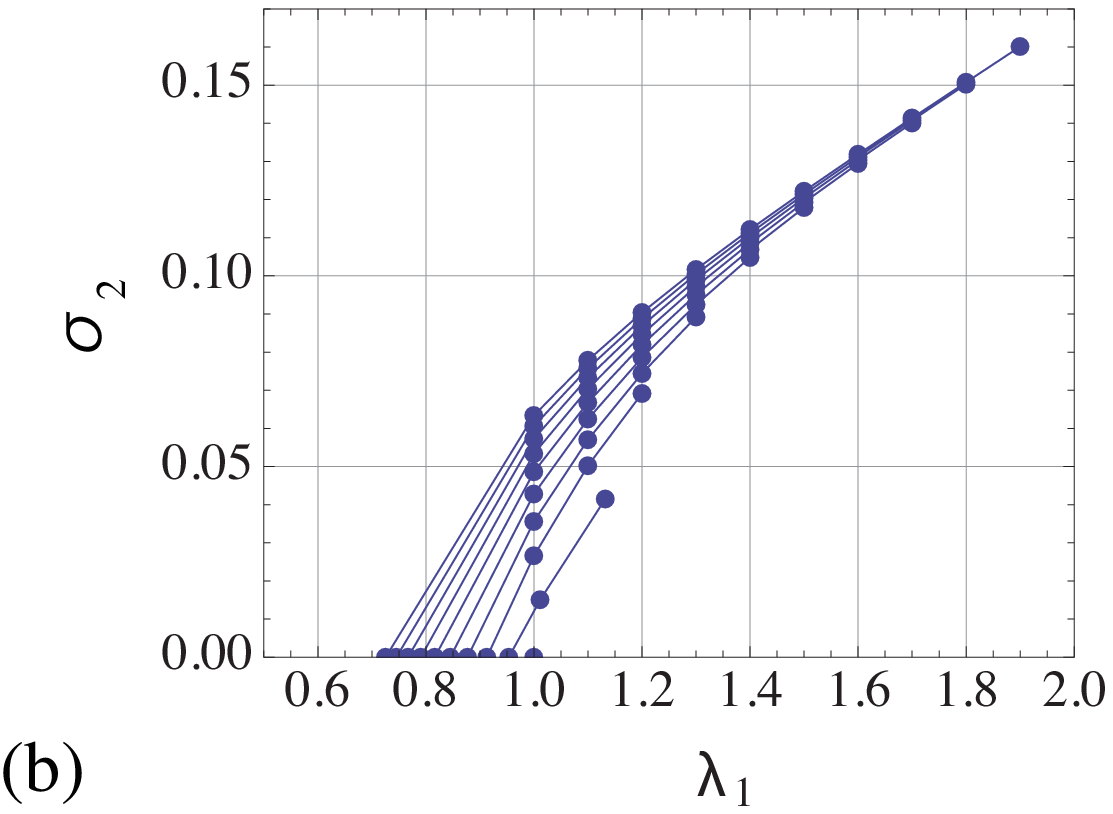}}
\end{tabular}
\caption{Nominal stresses along the $x$- and $y$-directions, (a) $\sigma_{1}$ and (b) $\sigma_{2}$, respectively, calculated from the new model under combinations of the deformation ratios $\lambda_1$ and $\lambda_2$.
Here the adjustable parameter $\gamma$ is set to $0.1$ and the parameter values $C, \, \alpha, \, \beta, \, \nu_1$ and $\nu_2$ in the case of $r_d=0.0$ and $N=256$ are used.
Each solid line stands for the $\lambda_2$-dependence of the stress under a fixed deformation $\lambda_1$.
The bottom line corresponds to $\lambda_1=1.1$, the top line corresponds to $\lambda_1=1.9$ and the intervals are $0.1$.
This graph shows the sum of the topological stress of our new model and the linear Hookean stress.
We determined the coefficient $C_{10}$ of the Hookean term in Eq.\ (\ref{5W}) so as to agree with the biaxial experimental data~\cite{U1}.
}
\label{fig11}
\end{center}
\end{figure}

Let us finally consider the sample in biaxial extension illustrated in Fig.\ \ref{fig10}.
The results for biaxial deformation are plotted in Fig.\ \ref{fig11} (a) and (b).
Our new model exhibits the stress-stain curves that are qualitatively consistent with an experimental result~\cite{U1}.
 According to the Ref.~\cite{U2}, there are few theoretical models that reproduce the biaxial experiment.
The present result for biaxial deformation is a remarkable advantage of our new model.

\section{Conclusions}

We have improved the GP model as follows.
We have proposed that the probability of entangled loops, $f(\zeta)$, must be, in principle, different from the linking probability of equilibrium SAPs, $P_{\mathrm{L}} (\zeta)$, since the loop topologies are determined by the nonequilibrium process of network forming, when the polymers are not distributed as SAPs.
Moreover, we have introduced a new phenomenological parameter $\gamma$ in order to evaluate $f(\zeta)$ such that $f(\zeta) = \gamma P_{\mathrm{L}} (\zeta)$.




The model qualitatively well represents nonlinear rubber elasticity both in extension and compression regime for $0.01 < \gamma < 0.1$.
Here we emphasize that our model also explains the biaxial experiments, while few models in the previous studies can reproduce them.
We thereby claim that the present model gives an essential step to quantitative analysis of the nonlinear rubber elasticity.


The primary advantage of the present model is that it has few adjustable parameters.
If the parameter $\gamma$ is given, the model stress is calculated by the structural and topological features of a network without any other adjustable parameters.
Further study is required to clarify the appropriate way of determining the parameter $\gamma$.

\section*{Acknowledgments}
The authors wish to express their gratitude to Prof. K. Urayama for the stimulating and helpful discussions.
We would like to thank Prof. T. Deguchi for drawing our attention to this subject.
We are also deeply indebted to Prof. N. Hatano for critical reading of the manuscript.
One of the authors (K.T.) was granted by KAKENHI (Grant-in-Aid for Scientific
Research) on Priority Area ``Soft Matter Physics'' from the Ministry of
Education, Culture, Sports, Science and Technology of Japan. 
Finally, we acknowledge the FUJUKAI foundation for their financial support.

\end{document}